\documentclass[aps,showpacs,preprintnumbers,amsmath,amssymb]{revtex4}
\usepackage{dcolumn}
\usepackage[dvips]{epsfig}

\begin{document}
\baselineskip=0.8 cm
\title{\bf $P-V$ criticality of AdS black hole in $f(R)$ gravity}

\author{Songbai Chen\footnote{csb3752@hunnu.edu.cn}, Xiaofang Liu, Changqing Liu
}

\affiliation{Institute of Physics and Department of Physics, Hunan
Normal University,  Changsha, Hunan 410081, People's Republic of
China \\ Key Laboratory of Low Dimensional Quantum Structures \\
and Quantum Control of Ministry of Education, Hunan Normal
University, Changsha, Hunan 410081, People's Republic of China}

\begin{abstract}
\baselineskip=0.6 cm
\begin{center}
{\bf Abstract}
\end{center}

We study thermodynamics of a charged AdS
black hole in the special $f(R)$ correction
 with the constant Ricci scalar curvature.
 Our results show that the the $f(R)$ correction
 influences the Gibbs free energy and the phase transition
 of system. The ratio $\rho_c$ occurred at the critical point
 increases monotonically with the derivative term $f'(R_0)$.
 We also disclose that the critical exponents are the same as
 those of the liquid-gas phase transition in the Van der Waals
 model, which does not depend on the $f(R)$ correction considered here.

\end{abstract}

 \maketitle
\newpage
\section{Introduction}

$f(R)$ gravity is a kind of important modified gravity theory, which
generalizes Einstein's gravity by adding higher powers of the scalar
curvature $R$, the Riemann and Ricci tensors, or their derivatives
in the usual lagrangian formulation \cite{fR1,fR2,fR21}. Without
unknown forms of dark energy or dark matter, $f(R)$ gravity can
explain the accelerating expansion of the current Universe. It can
also mimic the inflation and structure formation in the early
Universe. Thus, a lot of efforts have been focused in the study of
$f(R)$ gravity including not only the applications of $f(R)$
theories
 on gravitation and cosmology, but also many observational and
 experimental methods to distinguish them from general relativity \cite{fR2,fR21}.

It is of interest to extend the study of $f(R)$ gravity to the black
hole physics, which could provide us some features of black holes
differed from that obtained in Einstein's gravity. In general, the
black hole solution is difficult to be obtained in the $f(R)$ theory
since the field equations of $f(R)$ gravity become more complicated
than that in Einstein's gravity, even without the presence of a
matter field. With the condition that the energy-momentum tensor is
traceless for the matter field, Moon \textit{et al} \cite{fBH6}
obtain an exact analytical four-dimensional solution from $R+f(R)$
theory
 coupled to a matter field, which is later extended to the rotating
 cases \cite{fBH7}. The higher dimensional charged solution \cite{AS}
 are obtained only in the case of power-Maxwell field with $d=4p$, where $p$
  is the power of conformally invariant Maxwell lagrangian. Moreover,
  some other solutions of black hole in $f(R)$ gravity have been also
  constructed in \cite{fBH1,fBH2,fBH3,fBH4,fBH5}.

Thermodynamical properties of the AdS black hole has been a subject
of intense study for the past decades because that it is dual to a
thermal
 state on the conformal boundary in terms of the AdS/CFT correspondence.
 The duality has been recently applied to study the strongly correlated
 condensed matter physics from the
gravitational dual. Recently, Kubiznak \textit{et al} \cite{DKRB}
investigated the critical behavior of charged AdS black holes by
treating the cosmological constant as a thermodynamic pressure and
its conjugate quantity as a thermodynamic volume \cite{sq1}. It is
found that in the system of charged AdS black hole the small-large
black hole phase transition possesses the same critical behavior of
liquid-gas phase transitions in the Van der Waals model. The similar
critical behaviors are disclosed in the spacetimes of a rotating AdS
black hole \cite{SGRB} and a high dimensional
Reissner-Nordstr\"om-AdS black hole \cite{AM1}. The same qualitative
properties are also found in the AdS black hole spacetime with the
Born-Infield electrodynamics \cite{SGRB,BF1}, with the power-Maxwell
field \cite{SM1} and with Gauss-Bonnet correction \cite{Wei}.

The aim of this Letter is to study the thermodynamics of a charged
AdS black hole \cite{fBH6} in the special $f(R)$ correction in which
the Ricci scalar curvature remains the constant ($R=R_0$) and to
probe the effects of the $f(R)$ correction on thermodynamical
quantities and the small-large black hole phase transition. Finally,
we discuss the analogy of black hole with Van der Waals liquid-gas
system.

We firstly review the four-dimensional charged AdS black hole
obtained by Moon \textit{et al }\cite{fBH6} in the $R+f(R)$ gravity
with the constant Ricci scalar curvature. The action describing a
four-dimensional charged AdS black hole in the $R+f(R)$ gravity can
be expressed as \cite{fBH6}
\begin{eqnarray}
S=\int_{\mathcal{M}} d^{4}x\sqrt{-g}[R+f(R)-F_{\mu\nu}F^{\mu\nu}],
\label{action}
\end{eqnarray}
where $R$ is the Ricci scalar curvature and $f(R)$ is an arbitrary
function of $R$. $F_{\mu\nu}$ is electromagnetic field tensor which
is related to the electromagnetic potential $A_\mu$ by
$F_{\mu\nu}=\partial_{\mu}A_{\nu}-\partial_{\nu}\partial_{\mu}$.

Varying the action (\ref{action}), one can obtain the equations of
motion for gravitational field $g_{\mu\nu}$ and the gauge field
$A_{\mu}$
\begin{eqnarray}
R_{\mu\nu}[1+f'(R)]-\frac{1}{2}g_{\mu\nu}[R+f(R)]+(g_{\mu\nu}\nabla^2-
\nabla_{\mu}\nabla_{\nu})f'(R)=T_{\mu\nu}. \label{Esteq}
\end{eqnarray}
and
\begin{eqnarray}
\partial_{\mu}(\sqrt{-g}F^{\mu\nu})=0,
\label{Emeq}
\end{eqnarray}
respectively. In order to obtain an analytical solution of the
equation (\ref{Esteq}), Moon \textit{et al }\cite{fBH6} consider
only the case of the constant Ricci scalar curvature $R=R_0=const$
for the sake of convenience. In this simple case, Eq. (\ref{Esteq})
can be rewritten as
\begin{eqnarray}
R_{\mu\nu}[1+f'(R_0)]-\frac{g_{\mu\nu}}{4}R_0[1+f'(R_0)]
=T_{\mu\nu}, \label{Esteq1}
\end{eqnarray}
since the Maxwell energy-momentum tensor is traceless in the
four-dimensional spacetime. The general metric for a
four-dimensional charged static spherically symmetric black hole can
be expressed as
 \begin{eqnarray}
ds^2=-N(r)dt^2+\frac{dr^2}{N(r)}+r^2(d\theta^2+\sin^2\theta
d\phi^2).\label{metric}
\end{eqnarray}
Solving Eq. (\ref{Emeq}) in the four-dimensional static spacetime
(\ref{metric}), one can get the Maxwell field
\begin{eqnarray}
F_{tr}=\frac{q^2}{r^2}.\label{Emeq1}
\end{eqnarray}
Inserting Eqs.(\ref{metric}) and (\ref{Emeq1}) into
Eq.(\ref{Esteq1}), one can find that the metric (\ref{metric}) has
the form \cite{fBH6}
\begin{eqnarray}
N(r)=1-\frac{2m}{r}+\frac{q^2}{br^2}-\frac{R_0}{12}r^2.\label{metric1}
\end{eqnarray}
Here $b=[1+f'(R_0)]$. The parameters $m$ and $q$ are related to the
ADM mass $M$ and the electric charge $Q$ of the black hole by
\cite{AS}
\begin{eqnarray}
M=m b,~~~~~~~~~~Q=\frac{q}{\sqrt{b}},
\end{eqnarray}
and the electric potential $\Phi$ at horizon $r_+$ are
\begin{eqnarray}
\Phi=\frac{\sqrt{b}q}{r_+}.
\end{eqnarray}
Thus, the $f(R)$ correction modifies the mass $M$ and the electric
charge $Q$ of black hole and the corresponding thermodynamic
potentials. Treating the Ricci scalar curvature as
$R_0=-\frac{12}{l^2}=-4\Lambda$, one can find that the spacetime
described by solution (\ref{metric1}) is asymptotically AdS. In
other words, the Ricci scalar curvature $R_0$ is negative in the AdS
background. The Hawking temperature $T$ with the outer event horizon
$r=r_+$ is
\begin{eqnarray}
T=\frac{N'(r_+)}{4\pi}\bigg|_{r=r_+}=\frac{1}{4\pi
r_+}\bigg[1-\frac{q^2}{r_{+}^{2}b}
-\frac{R_0r_{+}^{2}}{4}\bigg],\label{Temp}
\end{eqnarray}
and the entropy of the black hole is \cite{fBH6,AS}
\begin{eqnarray}
S=\pi r_{+}^{2}b.\label{entropy}
\end{eqnarray}
For an AdS black hole, one can get both the differential and
integral formulas of the first law of thermodynamics by treating the
cosmological constant as a variable related to the thermodynamic
pressure \cite{sq1}. Motivated by this spirit, we here interpret
$R_0$ as a thermodynamic pressure $P$, which is given by
\begin{eqnarray}
P=-\frac{bR_0}{32\pi},\label{Pres}
\end{eqnarray}
and the corresponding volume $V$ is
\begin{eqnarray}
V=\frac{4\pi r^{3}_+}{3}. \label{Vo}
\end{eqnarray}
Thus, for the charged AdS black hole (\ref{metric}), the Smarr
formula  can be expressed as
\begin{eqnarray}
M=2TS+\Phi Q-2PV, \label{Sm}
\end{eqnarray}
which is consistent with the usual charged AdS black hole without
$f(R)$ correction. The differential form the first law of
thermodynamics becomes
\begin{eqnarray}
d\bigg(\frac{M}{b}\bigg)=Td\bigg(\frac{S}{b}\bigg)+\bigg(\frac{\Phi}{b}\bigg)
dQ+Vd\bigg(\frac{P}{b}\bigg). \label{dfen}
\end{eqnarray}
Obviously, it recovers the standard first law of thermodynamics for
the black hole as $b=1$.
 From Eqs. (\ref{Pres}) and (\ref{dfen}),
 we find that $R_0$ and $f'(R_0)$ modifies the thermodynamic pressure $P$
 of the system and the differential form the first law of thermodynamics.
 Thus, it is expected that the $f(R)$ correction will influence the thermodynamical
 phase transition of the AdS black hole and  could lead to some new features for the
 thermodynamics of black hole.

From the Hawking temperature (\ref{Temp}) and the pressure
(\ref{Pres}) of the system, one can get the equation of state
$P=P(T,V)$,
\begin{eqnarray}
P=\frac{bT}{2r_+}-\frac{b}{8\pi r_{+}^{2}}+\frac{q^2}{8\pi
r_{+}^{4}},\label{EOS}
\end{eqnarray}
for a fixed charge $q$. Here $r_+$ is a function of the
thermodynamic volume $V$. As doing in Refs.\cite{DKRB}, one can find
that the geometric quantities $P$ and $T$ can be translated into
physical pressure and temperature of system by using dimensional
analysis and $l^{2}_P=G\hbar/c^3$
\begin{eqnarray}
[\text{Press}]=\frac{\hbar
c}{l_{P}^2}P,~~~~~~~~~~~~[\text{Temp}]=\frac{\hbar c}{k}T.
\end{eqnarray}
And then the physical pressure and physical temperature are given by
\begin{eqnarray}
\text{Press}=\frac{\hbar c}{l_{P}^{2}}P=\frac{\hbar
c}{l_P^{2}}\frac{T}{2r_+} +\cdots=\frac{k
\text{Temp}}{2l_p^{2}}+\cdots.
\end{eqnarray}
In order to compare it with the Van der Waals equation
\cite{DKRB,SGRB}, we must identify the specific volume $v$ with the
horizon radius $r_+$ of the black hole by
\begin{eqnarray}
v=2r_{+}l_{P}^{2}.
\end{eqnarray}
After finishing these operations, one can find that the equation of
state (\ref{EOS}) can be rewritten as
\begin{eqnarray}
P=\frac{bT}{v}-\frac{b}{2\pi v^2}+\frac{2q^2}{\pi
v^{4}}.\label{EOS1}
\end{eqnarray}
\begin{figure}[ht]
\begin{center}
\includegraphics[width=5cm]{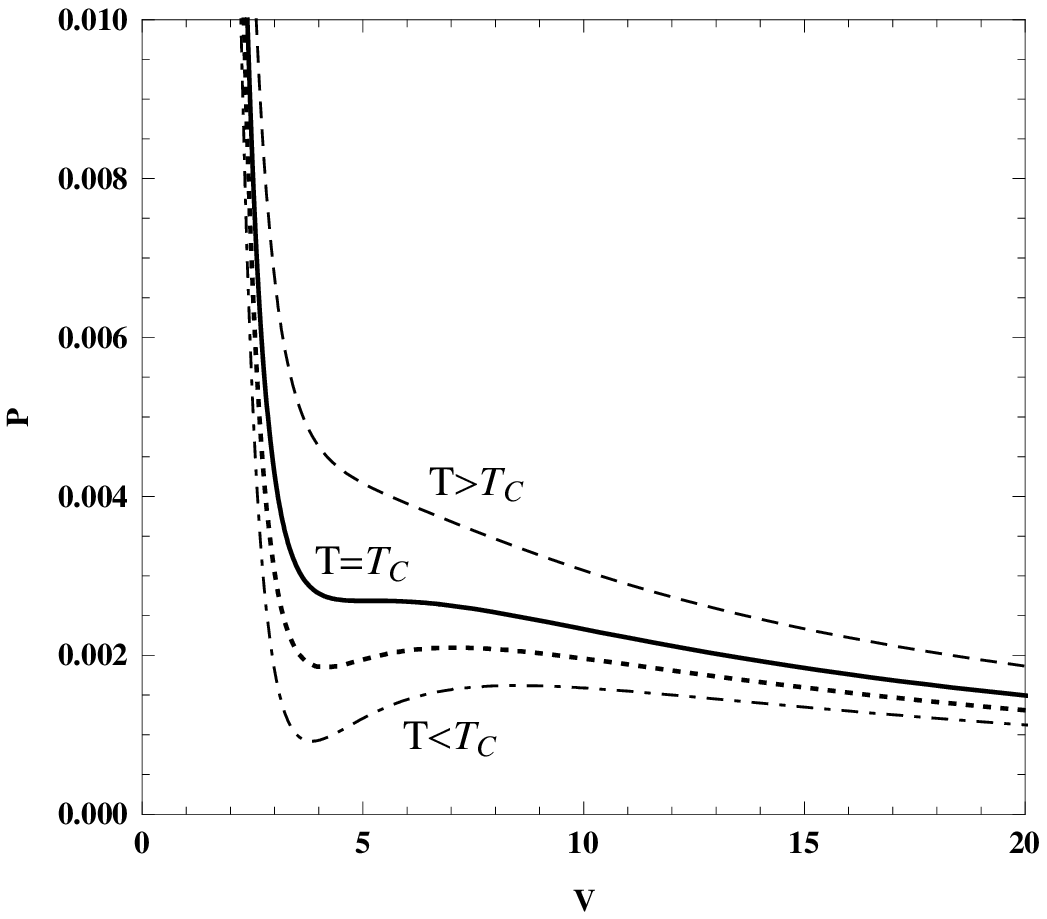}\includegraphics[width=5cm]{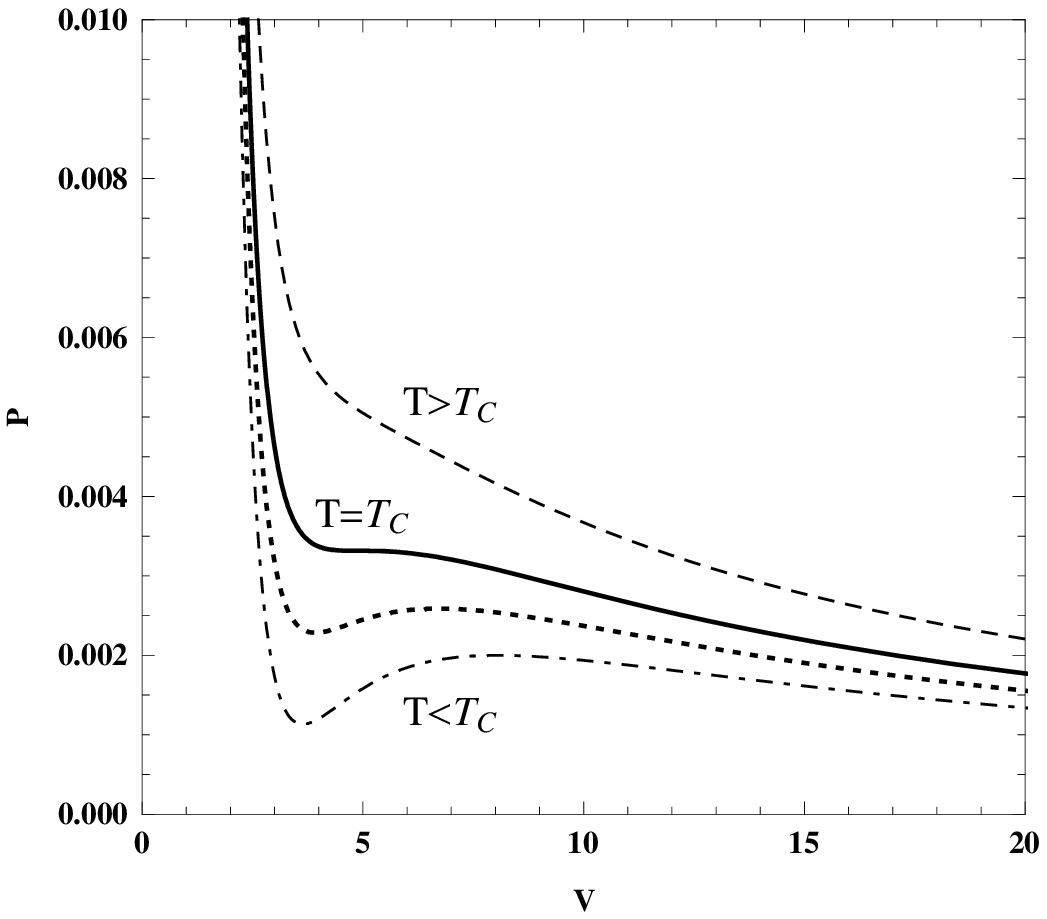}
\includegraphics[width=5cm]{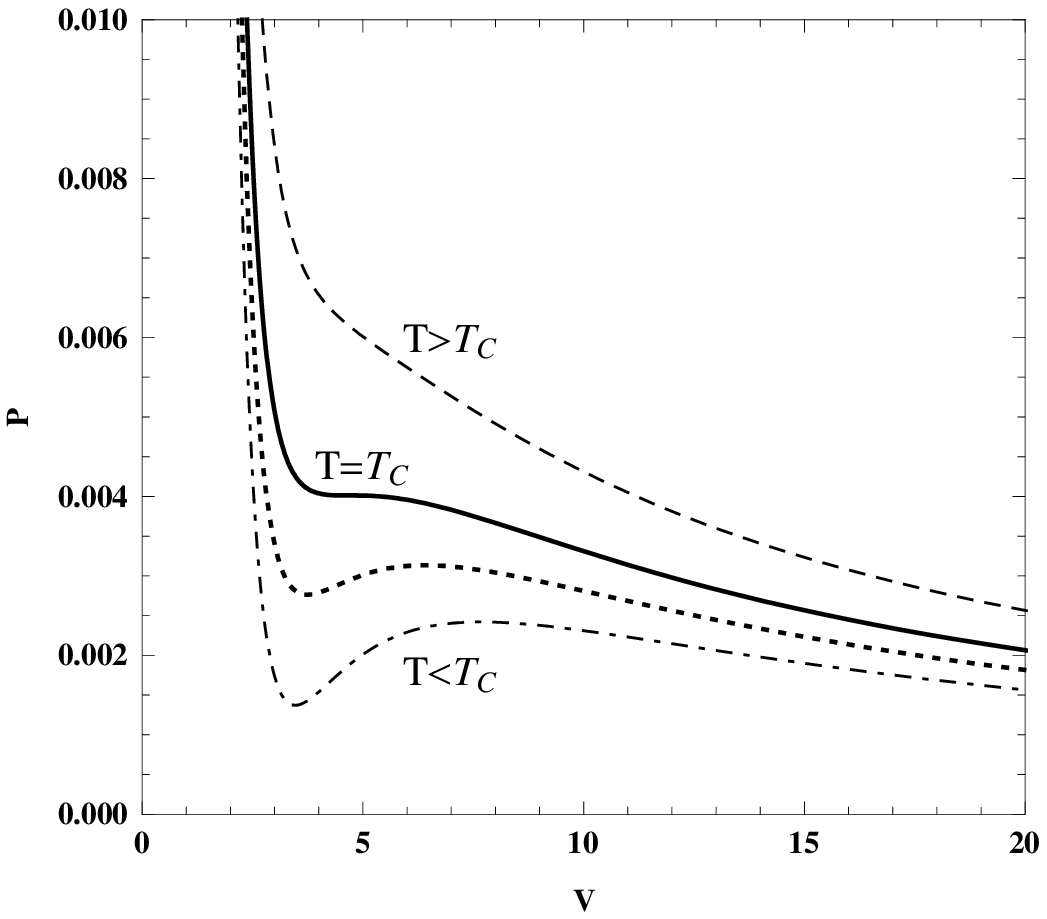}
\caption{$P-V$ diagram of a four-dimensional charged AdS black hole
in the $f(R)$ gravity. The upper dashed line correspond to the
¡°ideal gas¡± one-phase behavior for $T>T_c$, the critical isotherm
$T=T_c$is denoted by the thick solid line, lower dotted and
dot-dashed lines correspond to two-phase state occurring for
$T<T_c$. The left, middle and right plane are for $f'(R_0)=-0.1$,
$0$, and $0.1$, respectively.  Here we set $q=1$.}
\end{center}
\end{figure}
In Fig.(1) , we plot the $P-V$ diagram1 for a four-dimensional
charged AdS black hole in the $f(R)$ with the different $f'(R_0)$
for fixed $q=1$. Obviously, there exist a small-large black hole
phase transition in this system. Fig.(1) tells us that with increase
of $f'(R_0)$ the critical temperature $T_c$ increases.

The critical points in $P-V$ diagram can be obtained from
\begin{eqnarray}
\frac{\partial P}{\partial v}=0,\;\;\;\;\;\; \frac{\partial^2
P}{\partial v^2}=0,
\end{eqnarray}
which yields
\begin{eqnarray}
v_c=\frac{2q\sqrt{6}}{\sqrt{b}},~~~~~~~~~~
T_c=\frac{\sqrt{6b}}{18\pi q},~~~~~~~~~ P_c=\frac{b^2}{96\pi q^2}.
\end{eqnarray}
These formulas tell us that with increase of $f'(R_0)$ both of the
critical temperature $T_c$ and the pressure $P_c$ increase, but the
critical volume
 $v_c$ decreases, which is consistent with that shown in Fig. (1).
In order to ensure the occurrence of the phase transition, all of
the critical values $P_c$, $T_c$ and $v_c$ must be positive, which
implies
 that the condition $b=1+f'(R_0)>0$ should be satisfied for the black hole.

The ratio occurred at the critical point is given by
\begin{eqnarray}
\rho_c=\frac{P_c v_c}{T_c}=\frac{3b}{8},\label{rhoc}
\end{eqnarray}
for the charged AdS black hole in the $f(R)$ gravity (\ref{metric}).
Clearly, the $f(R)$ correction changes the ratio $\rho_c$ occurred
at the critical point. It depends on the concrete form of $f(R)$
gravity.
 With increase of $f'(R_0)$, the ratio $\rho_c$ increases monotonically.
  When the derivative term $f'(R_0)$
disappears, one can find that the ratio $\rho_c$ is recovered to
that in the usual four-dimensional charged AdS black hole spacetime.
From Eq.(\ref{rhoc}), one can find that the charged AdS black hole
in the $f(R)$ gravity possesses the similar feature of the Van
der-Waals fluid.

Let us now analyze the Gibbs free energy of the system. For a
canonical ensemble, the Gibbs free energy is $G=M-TS$. This yields
that for a charged AdS black hole (\ref{metric1}) in the $f(R)$
gravity the Gibbs free energy is
\begin{eqnarray}
G=\frac{1}{4}\bigg[br_{+}-\frac{8\pi
Pr_{+}^{3}}{3}+\frac{3q^{2}}{r_+} \bigg].
\end{eqnarray}
\begin{figure}[ht]
\begin{center}
\includegraphics[width=5cm]{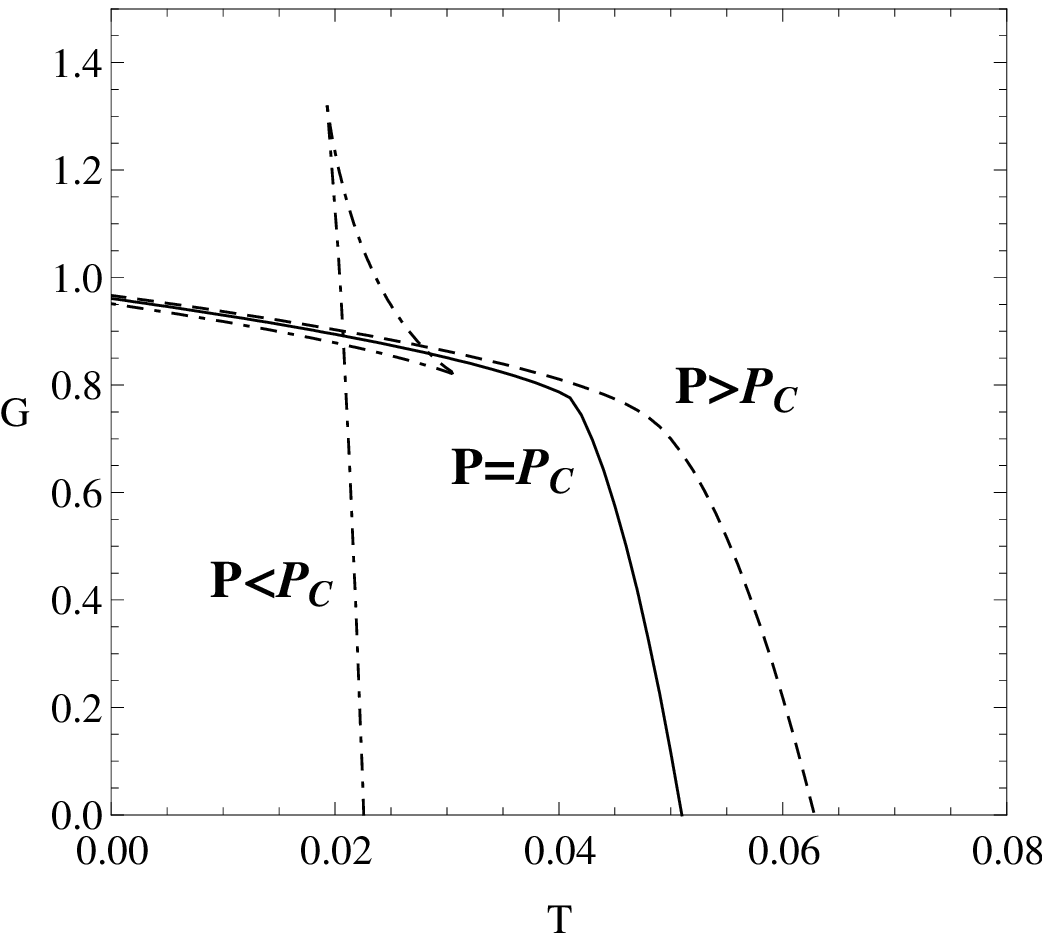}\includegraphics[width=5cm]{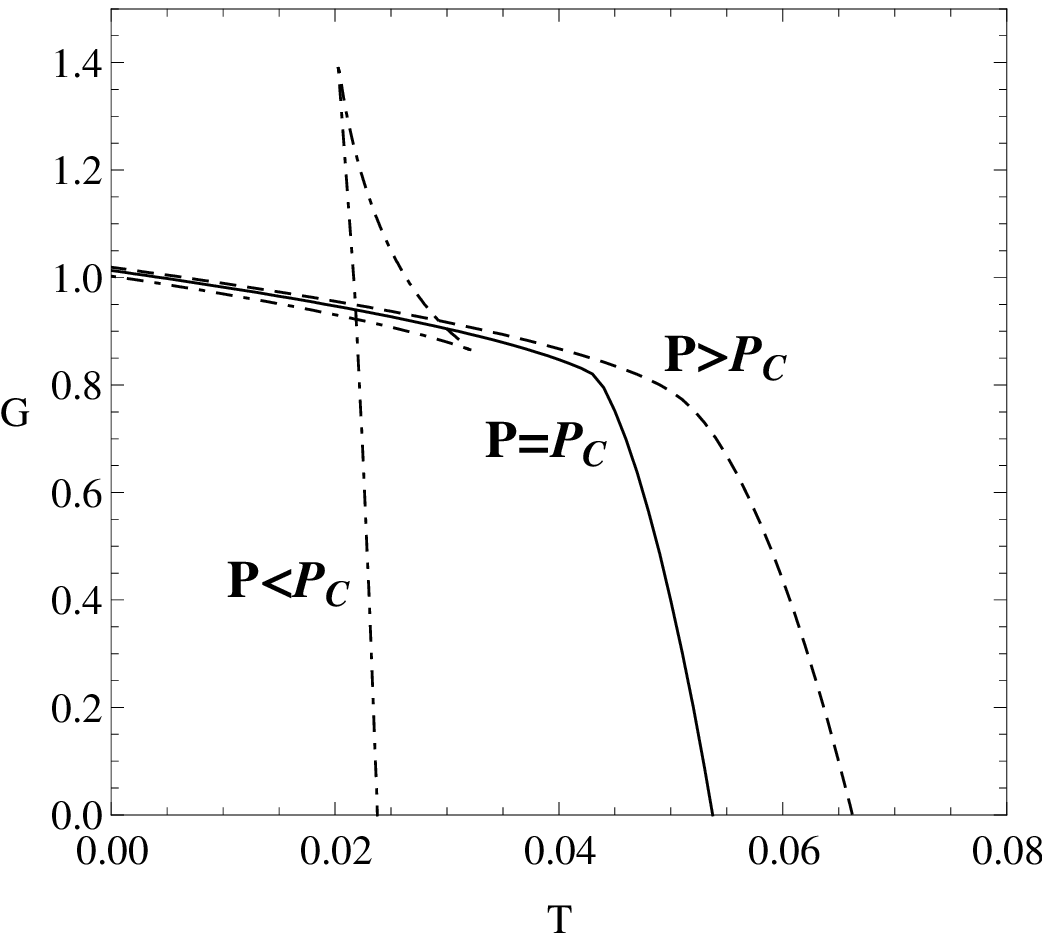}
\includegraphics[width=5cm]{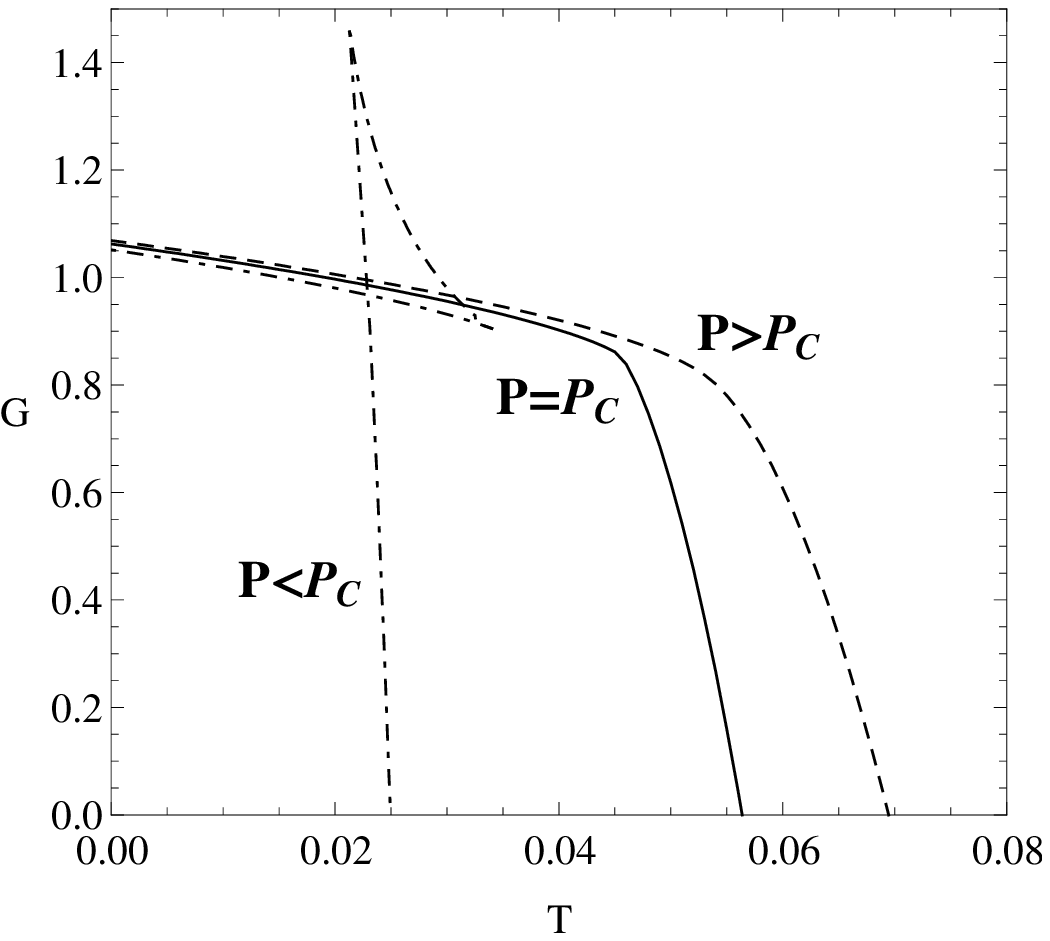}
\caption{Gibbs free energy  of a four-dimensional charged AdS black
hole in the $f(R)$ gravity. The dashed, solid, and dot-dashed lines
correspond to the cases $P/P_c=1.5$, $1$, and $0.2$. The left,
middle and right panels
 are for $f'(R_0)=-0.1$, $0$, and $0.1$, respectively.  Here we set $q=1$.}
\end{center}
\end{figure}
We plot the change of the free energy $G$ with $T$ for fixed $q$ in
Fig.(2). The presence of the characteristic ``swallow tail" behavior
of the free energy
 $G$ means that the small-large black hole transition occurred in the system is a
first order phase transition. The two-phase existence is also shown
in Fig. (3),
 in which the coexistence line is obtained by a fact that two phases share the
 same  Gibbs free energy and temperature during the phase transition.
 The coexistence line can be also produced by Maxwell's equal area law
 and the Clausius-Clapeyron equation. From Fig. (3), one can find that
 the coexistence line depends on the form of $f(R)$ and its slope
 increases with the derivative term $f'(R_0)$.
\begin{figure}[ht]
\begin{center}
\includegraphics[width=6.3cm]{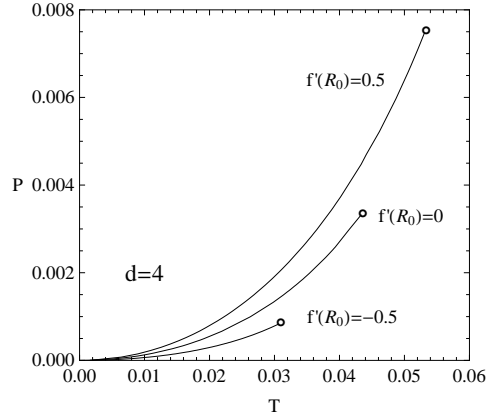}
\caption{Coexistence line of a four-dimensional charged AdS black
hole in $f(R)$ gravity. The critical point is shown by a small
circle at the end of the coexistence line. The left, middle and
right
 panels are for $f'(R_0)=-0.5$, $0$, and $0.5$, respectively.  Here we set $q=1$.}
\end{center}
\end{figure}

The critical exponents are very important for the phase transition
occurred in the thermodynamic system. Here, we will discuss the
critical exponents $\alpha, \beta, \gamma, \delta$ for the charged
AdS black hole in the $f(R)$ gravity. Similarly,  the entropy
(\ref{entropy}) $S$  can be expressed as a function of $T$ and $V$,
\begin{eqnarray}
S=S(T,V)=b\bigg(\frac{3V}{4\pi}\bigg)^{2/3}.
\end{eqnarray}
Obviously, it is independent of the temperature $T$, which yields
directly that the specific heat vanishes $C_V=0$ and the critical
exponent $\alpha$ is zero. Adopting to the reduced thermodynamics
\begin{eqnarray}
p=\frac{P}{P_c},\;\;\;\nu=\frac{v}{v_c},\;\;\;\tau =\frac{T}{T_c},
\end{eqnarray}
one can find that the equation of state (\ref{EOS}) can be rewritten
as
\begin{eqnarray}
p=\frac{8\tau}{3\nu} -\frac{2}{\nu^2}+\frac{1}{3\nu^{4}}.\label{p1}
\end{eqnarray}
It is the same as that of a charged AdS black hole without $f(R)$
correction, which means that the $f(R)$ correction with the constant
Ricci scalar curvature does not change the critical exponents
$\alpha, \beta, \gamma, \delta$ for the system of black hole. In
other words, the critical exponents for a charged AdS black hole in
$f(R)$ gravity with the constant Ricci scalar curvature coincide
with those of the Van der Waals fluid and those of in the mean field
theory.

In this Letter, we consider thermodynamics of a charged AdS black
hole
 in the special $f(R)$ correction in which the Ricci scalar curvature
  is the constant $R=R_0$. And then we study the effects of the $f(R)$
  correction on thermodynamical quantities and the phase transition.
  Our results show that the $f(R)$ correction changes the equation of
   state of the system and influence the critical temperature $T_c$,
    the volume $v_c$,  and the pressure $P_c$. With increase of $f'(R_0)$,
    both of the critical temperature $T_c$ and the pressure $P_c$ increase,
    but the critical volume $v_c$ decreases. Moreover, we also find the
    ratio $\rho_c$ occurred at the critical point depends on the form of
    $f(R)$ and it increases monotonically with the derivative term $f'(R_0)$.
    As the derivative term $f'(R_0)$ disappears, it is recovered to that in
    the usual four-dimensional charged AdS black hole spacetime without $f(R)$
    correction. Moreover, we also find that
the $f(R_0)$ correction affects the Gibbs free energy and the slope
of the two-phase coexistence line in the $P-T$ plane. However, the
critical exponents $\alpha, \beta, \gamma, \delta$ are independent
of the special $f(R)$ correction and coincide with those of the Van
der Waals fluid and of in the mean field theory, which could be the
universal property for such kinds of phase transition.

\begin{acknowledgments}
This work was  partially supported by the National Natural Science
Foundation of China under Grant No.11275065, the NCET under Grant
No.10-0165, the PCSIRT under Grant No. IRT0964,  the Hunan
Provincial Natural Science Foundation of China (11JJ7001) and the
construct program of key disciplines in Hunan Province.
\end{acknowledgments}


\vspace*{0.2cm}
\begin{thebibliography}{99}

\baselineskip=0.6 cm



\bibitem{fR1}Buchdahl H A 1970 Mon. Not. R. Astron. Soc {\bf150}  1

Starobinsky  A A 1980  Phys. Lett. B {\bf91} 99

\bibitem{fR2} Felice A D  and Tsujikawa S 2010 Living Rev. Rel. {\bf13} 3

Capozziello S and Laurentis M D 2011 arXiv:1108.6266v2[gr-qc]

\bibitem{fR21} Nojiri S and Odintsov S D 2011  Phys. Rept. {\bf505} 59

\bibitem{fBH6} Moon T, Myung Y S and Son E J 2011 arXiv:1101.1153v2[gr-qc]

\bibitem{fBH7} Larranaga A 2011 arXiv:1108.6325v1[gr-qc]

Cembranos J A R, Cruz-Dombriz A D L and Romero P J 2011
arXiv:1109.4519v2[gr-qc]


\bibitem{AS} Sheykhi A 2012 Phys. Rev. D {\bf86} 024013


\bibitem{fBH1} Cognola G, Elizalde E, Nojiri S, Odintsov S D and Zerbini S 2005 JCAP {\bf0502} 010

\bibitem{fBH2} Sebastiani L and Zerbini S, arXiv:1012.5230v3[gr-qc]

\bibitem{fBH3} Cruz-Dombriz , Dobado A and  Maroto A L 2009 Phys. Rev. D {\bf80} 124011

    Cembranos J¡¡A R, Cruz-Dombriz  A D L and Romero P J 2012 arXiv:1202.0853v1[gr-qc]


\bibitem{fBH4}Hendi S H 2010 Phys. Lett. B {\bf690}, 220

Hendi S H  and Momeni D 2011 Eur. Phys. J. C {\bf71} 1823

\bibitem{fBH5}Olmo G J and Garcia D R 2011 Phys. Rev. D {\bf84} 124059
 Mazharimousavi S H and Halilsoy M 2011  Phys. Rev. D {\bf84} 064032

\bibitem{DKRB} Kubiznak D and Mann R B 2012 JHEP {\bf07} 033

\bibitem{sq1} Wang S, Wu S Q, Xie F and Dan L 2006 Chin. Phys. Lett.
{\bf23} 1096
 Sekiwa Y 2006 Phys. Rev. D {\bf73} 084009

Banados M, Teitelboim C and Zanelli J 1992 Phys. Rev. Lett. {\bf69}
1849

Banados M, Henneaux M, Teitelboim C and Zanelli J 1993 Phys. Rev. D
{\bf48} 1506

Wu S Q 2005 Phys. Lett. B {\bf608} 251

Shankaranarayanan S 2003 Phys. Rev. D {\bf67} 084026

Cai R G 2002 Phys. Lett. B {\bf525} 331


Gibbons G W, Lu H, Page D N  andPope C N 2005 J. Geom. Phys. {\bf53}
49


Kastor D,  Ray S and Traschen J 2009 Class. Quant. Grav. {\bf26}
195011


Kastor D,  Ray S and Traschen J 2010 Class. Quant. Grav. {\bf27}
235014


\bibitem{SGRB} Gunasekaran S and  Mann R B,2012 JHEP {\bf11} 110

\bibitem{AM1} Belhaj A,  Chabab M, Moumni H E and Sedra M B 2012 Chin. Phys. Lett. {\bf29} 100401

\bibitem{BF1} Banerjee R and Roychowdhury D 2012 Phys. Rev. D {\bf85} 104043

Banerjee R and Roychowdhury D 2012 Phys. Rev. D {\bf85} 044040


Majhi B R and Roychowdhury D 2012 Class. Quant. Grav. {\bf29} 245012





\bibitem{SM1} Hendi S H and Vahidinia M  H 2012 arXiv:1212.6128v2 [hep-th]
\bibitem{Wei} Wei S W and Liu Y X 2013 Phys. Rev. D {\bf87} 044014




\end{thebibliography}
\end{document}